\documentstyle[12pt,clustering,psfig]{article}
\def\be{\begin{equation}}

\def\ee{\end{equation}}

\def\h3Mpc{h^{-3}{\rm Mpc}^3}

\def\h3Mpcinv{h^{3}{\rm Mpc}^{-3}}

\def\spose#1{\hbox to 0pt{#1\hss}}
\def\simlt{\mathrel{\spose{\lower 3pt\hbox{$\mathchar"218$}}
     \raise 2.0pt\hbox{$\mathchar"13C$}}}
\def\simgt{\mathrel{\spose{\lower 3pt\hbox{$\mathchar"218$}}
     \raise 2.0pt\hbox{$\mathchar"13E$}}}

\begin{document}
\heading{WEAK LENSING AS A PROBE OF DARK MATTER\footnote{To be published in the proceeding of the XXXIst Rencontres de Moriond, Les Arcs, France, January 20-27 1996}{}}

\centerline{L. Van Waerbeke$^{(1)}$, Y. Mellier$^{(2)}$}
{\it
\centerline{$^{(1)}$ Observatoire Midi Pyr\'en\'ees,}
\centerline{14 Av. Edouard Belin,}
\centerline{31400 Toulouse, France}
\centerline{$^{(2)}$Institut d'Astrophysique de Paris,}
\centerline{98$^{bis}$ Boulevard Arago,}
\centerline{75014 Paris, France}
}




\begin{abstract}{\baselineskip 0.4cm 
Weak gravitational lensing of distant galaxies can probe the
total projected mass distribution of foreground gravitational
structures on all scales and  has been used successfully to map 
the projected mass distribution of  rich intermediate redshift
clusters. This paper reviews the general concepts of the lensing analysis. 
We focus on the relation between the observable (shapes and fluxes)
and physical (mass, redshift) quantities
 and discuss some observational issues and recent developments on data analysis
which appear promising for a better measurement of the lensing
signatures (distortion and magnification) at very large scales.
}
\end{abstract}

\section{Introduction}

The dark matter (DM) component of gravitational structures is extensively
studied from the dynamical
analysis of the luminous component. Popular examples are the rotation
curves of galaxies, motions of galaxies in groups or in clusters, or large scale
velocity fields  from which the mass and the distribution of the DM   
 can be inferred, provided one assumes a dynamical state
(Virial state) and a geometry (sphericity) of the 
 gravitational system.  Unfortunately, in most cases these
hypothese are not fulfilled: for example, the Virial hypothesis applied to
clusters may be wrong, because clusters may be young gravitational 
objects.  Their mass profile could be alternatively obtained from 
 the  X-ray Bremsstralung emission of their intra-cluster gas 
  which depends on their total
mass distribution and their equilibrium state as well. Again, 
one has to assume a geometry and a thermodynamical state for the gas 
of photons and electrons.
Despite these difficulties, all these studies provide similar
trends, with the mass to light ratio M/L increasing with scale. For a typical
galaxy M/L ranges between $10-30$, but is roughly $10$ times larger for a
cluster
from both Virial and X-ray studies, leading to $\Omega\sim 0.02$ for the 
former and $\Omega\sim 0.2$ for the later.
This means that the mass cannot be concentrated only within the 
central visible parts of galaxies.

Gravitational lensing provides a direct measurement of the projected mass
density without additional hypothesis on the dynamical state or on the geometry 
of the mass distribution.
Provided that we can measure the optical distortion of background
objects caused by a foreground mass, it is possible to
constrain the projected mass distribution of this deflector. Recent
results of the lensing analysis on some clusters are summerized in
Table (1).

\vskip 1cm
\psfig{figure=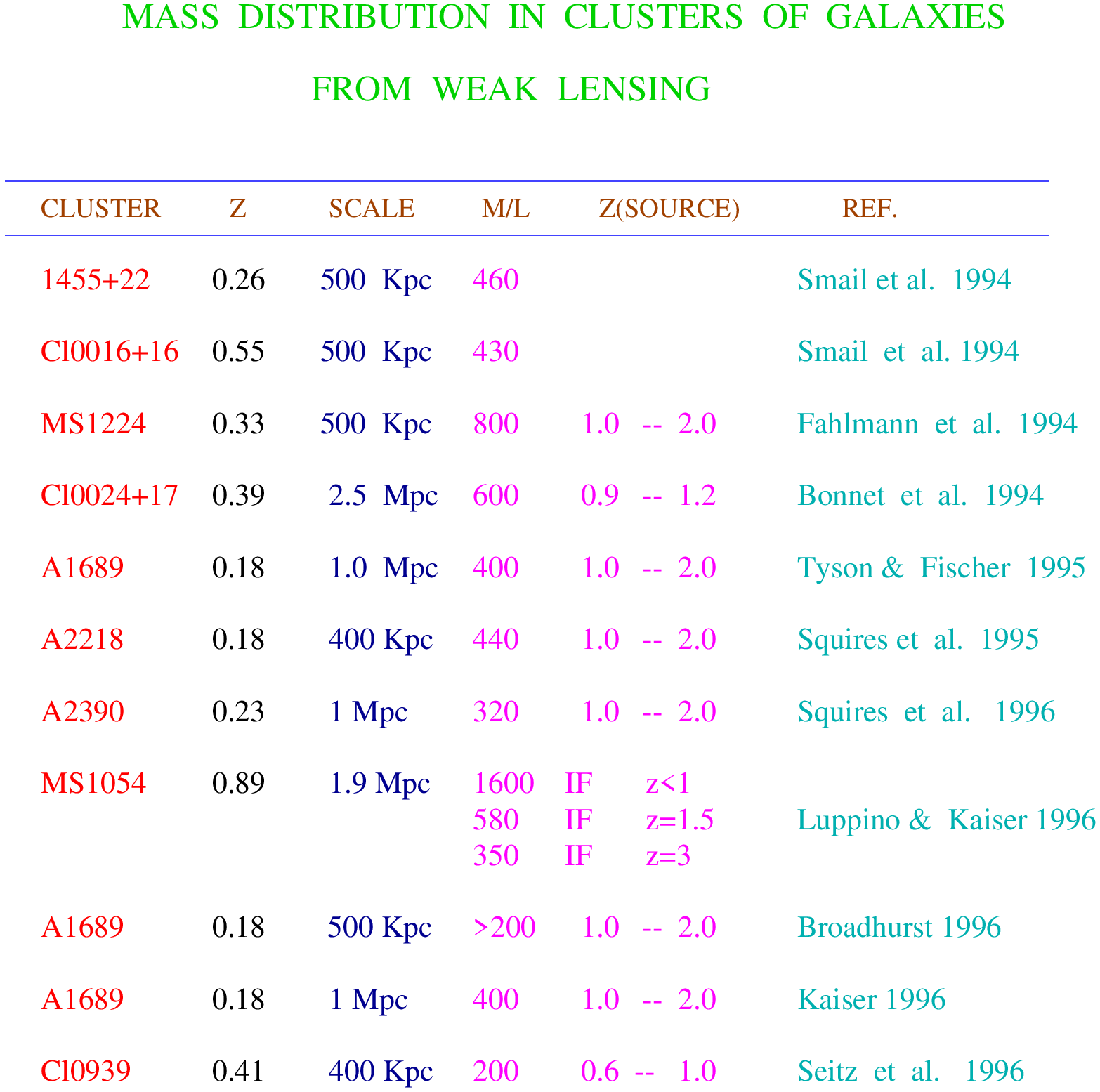,width=14. cm}
\vskip 0.5cm
\centerline{Table~1: Status of the observations.}
\vskip 1cm

These results are larger than the usual Virial or X-Ray analysis by a
factor 2 or 3.
Whether this discrepancy may be explained or not is not clear yet
(Miralda-Escud\'e \& Babul (1995), Navarro et al. (1995)),
and we do not discuss this here.
We discuss in this paper the general
method of the weak lensing analysis leading to these results.

\section{The weak lensing analysis}
\subsection{Basics of gravitational lensing}
Because of gravitational lensing, ray-lights are bended and the apparent
position $\vec \theta_I$ differs from the source position in absence of
lensing $\vec \theta_S$ by the quantity $\vec \alpha$:

\be
\vec \theta_I=\vec \theta_S+\vec \alpha
\ee

where $\vec \alpha$ is the gradient of the two-dimensional (projected on 
the line of sight) gravitational
potential $\phi$. The gravitational distortion of background objects is
described by the Jacobian of the transformation, namely the amplification
matrix $\cal A$ between the source and the image plane (for more
details, see Schneider, Ehlers \& Falco 1992):

\be
{\cal A}=\pmatrix{ 1-\kappa-\gamma_1 & -\gamma_2 \cr
-\gamma_2 & 1-\kappa+\gamma_1 \cr }
\ee

where $\kappa$ is the convergence, $\gamma_1$ and $\gamma_2$ are the
shear components, and are related to the Newtonian gravitational potential
$\phi$ by:

\be
\kappa={1\over 2} \nabla^2 \phi={\Sigma\over \Sigma_{crit}}
\ee
\be
\gamma_1={1\over 2}(\phi_{,11}-\phi_{,22}) \ ; \ \ \
\gamma_2=\phi_{,12} \ ,
\ee

where $\Sigma$ is the projected mass density and $\Sigma_{crit}$ is the
critical mass density which depends on the angular diameter distances $D_{ij}$,
($i,j=(o(bserver),l(ens),s(ource))$) involved in the lens configuration:

\be
\Sigma_{crit}={c^2\over 4\pi G}{D_{os}\over D_{ol} D_{ls}}
\ee

The quantities $\kappa$, $\gamma_1$, and $\gamma_2$ are not observables.
Only the magnification $\mu$ and the distortion $\delta$ are in
principle observable quantities because they are related to the fluxes
and the shapes of the objects (Miralda-Escud\'e 1991, Schneider \& Seitz
1995):

\be
\mu={1\over |{\cal A}|}={1\over (1-\kappa)^2-\gamma^2}
\ee
\be
\delta_i={2g_i\over 1+|g|^2} \ ; \ g_i={\gamma_i\over 1-\kappa}
\ee
Since we are interested in the large scale distribution of the Dark Matter ($>
0.5 Mpc$) we only focus the analysis on the weak lensing regime where $(\kappa,
\gamma) <<1$. The relations between the physical ($\gamma, \kappa$) and
observable ($\delta, \mu$) quantities become more simple:

\be
\mu=1+2\kappa
\ee
\be
\delta_i=2\gamma_i
\ee

The projected mass density $\Sigma$ of the lens is available from the
amplification $\mu$ using Eq. (8) and Eq. (3), or equivalently from the
distortion field by using Eq.(9) and the integration of Eq. (4)
(Kaiser, Squires (1993)):

\be
\kappa(\vec \theta_I)={-2\over \pi} \int d{\vec \theta} {\chi_i(\vec
\theta-\vec \theta_I) \gamma_i(\vec \theta_I)\over (\vec \theta-\vec
\theta_I)^2} +\kappa_0 \ ; \ \ \ \vec \chi(\vec \theta)=({\theta_1^2-
\theta_2^2\over \theta^2} \ , \  {2\theta_1 \theta_2 \over \theta^2})
\ee

$\kappa_0$ is the integration constant. In the weak lensing regime,
Eqs.(8) and (9) provide two
independant methods to map the total projected mass, using the
distortion of the background objects and the magnification of the
background objects.

\subsection{How observable quantities are measured?}
The gravitational distortion is not visible on a single galaxy in the weak
lensing regime because $\gamma<<{\bar \epsilon}$, where ${\bar \epsilon}$ is the
mean ellipticity of the galaxies. Fortunately a gravitational shear in a given
area of the sky distorts all the background galaxies by a same amount, and
the distortion can be measured from the mean polarization of these
galaxies. The distortion $\delta_i$ is computed from the shape of the
galaxies in the image plane. Each galaxy is assumed to be elliptical with an
ellipticity $\epsilon$ and an orientation $\theta$, and is
described by a polarization vector $\epsilon e^{2 i\theta}$ in a complex
formalism. The distortion is given by the sum of the polarization
vectors in a given area of the sky. No
information is required in the source plane, only the isotropy of the
orientation of the galaxies in the source plane is assumed.
A detailled description on the optimum detection and measurement of 
the shape of the galaxies
is given in Bonnet \& Mellier (1995) and Kaiser et al. (1995). 
This method has been
succesfully applied to several clusters (See Table(1)),
and to simulations to get the distortion.
However, the intrinsic ellipticity of the galaxies is a source
of noise, and the contribution of the random orientations of $N_g$ galaxies
to the value of the shear $\gamma_i$ is given by ${\bar \epsilon}/\sqrt{N_g}$.
An estimate of the mass using Eq. (10) requires an
estimation of $\Sigma_{crit}$, which implies the redshift of the
sources which is poorly known. Though this is not a critical issue for
nearby clusters ($z_l<0.2$) because ${D_{os}/ D_{ls}}\simeq 1$, it could
lead to a large uncertainty of the mass for more distant clusters (See
Table (1)).

Unfortunately it is impossible to get a true value for the mass only from
the shear map, even if we know the redshift of the sources,  because
a constant mass plane does not induce
any shear on background galaxies. Mathematically, this corresponds to the
unknown intregration constant $\kappa_0$ in Eq.(10).
This degeneracy may be broken if one    
measures the magnification $\mu$ which depends on the mass quantity inside
the light beam (Eq.(3)). While the
shear measurement does not require any information in the source plane, the
magnification measurement needs the observation of a reference (unlensed)
field to calibrate the magnification. Broadhurst et al. 1995 proposed to
compare the number count
$N(m,z)$ and/or $N(m)$ in a lensed and an unlensed field to measure
$\mu$. Depending on
the value of the slope $S$ of the number count in the reference field, we
observe a bias (more objects) or an anti-bias (less objects) in the
lensed field. The particular value $S=0.4$ corresponds to the case where
the magnification of faint objects is exactely compensated by the
dilution of the number count. This method was applied successfully on
the cluster A1689 (Broadhurst, 1995), but the signal to noise of the
detection remains 5 times lower than with the distortion method for a given
number of galaxies. The magnification may also be determined
by the changes of the image sizes at fixed surface brightness
(Bartelmann \& Narayan 1995).

The weakness of these methods is that
they require to measure the shape, size and magnitude of very faint
objects up to B=28, and this is not sure whether the measurement is optimum,
and whether systematic effects are avoided.
The determination of the shape parameter depends 
on the threshold level and the
convolution mask, and in any cases the information contained in pixels
fainter than the threshold level is lost.
Furthermore,  the measurement of the shape from the second
moment matrix is equivalent to the assumption that the objects are
elliptical, which is not true.
These remarks lead us to propose a new and independent method to analyse
the lensing effects, based on the auto-correlation function of the pixels in
CCD images, which avoids shape parameter measurements (Van
Waerbeke et al. 1996).

\section{The Auto-correlation method}
\subsection{Principle}
The CCD image is viewed as a density field rather than an image containing
delimited objects. The surface brightness in the image plane in the direction
$\vec \theta$ is related to the surface brightness in the source plane
$I^{(s)}$ by the relation:

\be
I(\vec \theta)=I^{(s)}({\cal A}\vec \theta)
\ee

and for the auto-correlation function (ACF):

\be
\xi(\vec \theta)=\xi^{(s)}({\cal A}\vec \theta)
\ee

To understand the meaning of this equation, let us write it in the weak lensing regime:

\be
\xi(\vec \theta)=\xi^{(s)}(\theta)-\theta \ \partial_{\theta} \xi^{(s)}(\theta)
[1-{\cal A}]
\ee

$\xi(\vec \theta)$ is the sum of an isotropic unlensed term
$\xi^{(s)}(\theta)$, an isotropic lens 
term which depends on $\kappa$, and an anisotropic term which depends on
$\gamma_i$.

Let us analyse which gravitational lensing information can be extracted
from the shape matrix $\cal M$ of $\xi$:

\be
{\cal M}_{ij}={\int d^2\theta \xi (\vec \theta) \theta_i \theta_j\over \int
d^2\theta \xi (\vec \theta)}
\ee

The shape matrix in the image plane is simply related to the shape
matrix in the source plane ${\cal M}^{(s)}$ by ${\cal M}_{ij}={\cal
A}_{ik}^{-1} {\cal A}_{jl}^{-1} {\cal M}^{(s)}_{kl}$. If the galaxies
are isotropically distributed in the source plane, 
$\xi^{(s)}$ is isotropic, and in that case ${\cal
M}^{(s)}_{ij}=M\delta_{ij}$, where $\delta_{ij}$ is the identity matrix.
Using the expression of the amplification matrix $\cal A$ we get the
general form for $\cal M$:

\be
{\cal M}={M(a+|g|^2)\over (1-\kappa)^2(1-|g|^2)} \pmatrix{
1+\delta_1 & \delta_2 \cr \delta_2 & 1-\delta_1 \cr }
\ee

The observable quantities (distortion $\delta_i$ and magnification
$\mu$) are given in terms of the components of the shape matrix:

\be
\delta_1={{\cal M}_{11}-{\cal M}_{22}\over tr{\cal M}} \ ; \ \ \
\delta_2={2{\cal M}_{12}\over tr{\cal M}} \ ; \ \ \
\mu=\sqrt{det{\cal M}\over M}
\ee

where $tr{\cal M}$ is the trace of $\cal M$. As for lensed galaxies, 
  we see that the
distortion is available from a direct measurement in the image plane while
the magnification measurement requires to know the value of $M$ which
is related to the light distribution in the source plane, or in an
unlensed reference plane. The ACF provides a new and independent 
way to measure
$\delta_i$ and $\mu$ which does not require shape, size or photometry
of individual galaxies. In the following we only describe the
measurement of the distortion using this method.
The case of the magnification which
requires an analysis of the sources galaxies in a reference field, will be
developped in a future work.

\subsection{The practical method}
By definition, the value of the ACF at a pixel position
$ij$ is $E_{ij}={1\over N_{pix}} \sum_{kl}(I_{i+k j+l}-\bar I)(I_{kl}-\bar I)$,
where $I_{kl}$ is the value of the pixel $kl$, $\bar I$ is the mean value
of the image, and $N_{pix}$ the number of terms in the sum. The ACF is computed
in a part of the image (a superpixel) where the shear
is assumed to be constant in intensity and direction.
Two strategies are possible to compute the ACF. First we
can remove all the unwanted objects (stars, bright galaxies, dead CCD
lines, cosmetic defaults,...) and compute the ACF from the rest of the
image. The main interest of this approach is that it works at the noise
level and even ultra faint objects are taken into account. The second
approach consists in selecting objects from a given criteria (magnitude,
colors, redshift,...), in surrounding them by a large circle, put the rest of
the image to zero and compute the ACF of the image containing these circles.

As for the case of individual galaxy we need to compute the shape matrix
of the ACF in an annular filter (Bonnet \& Mellier, 1995) to avoid the center,
where the signal is strongly polluted by the Point Spread Function (PSF),
and the external part, which is  dominated by the noise. The effects of 
the  PSF and the filter are calibrated by using simulations.

\subsection{Sources of errors}

The galaxies have not the same flux, size and profile and, 
by definition of the ACF,  are weighted by the square of their
flux. Since  this could change the statistical properties of the ACF,
it is better to work with selected objects by using the
second strategy of the ACF method. The idea is to weight each circle which
contains an object by a  multiplicative term defined as $[{1\over N_{pix}}
\sum_{ij} (I_{ij}-\bar I)^2]^{-1/2}$, where
$N_{pix}$ is the number of pixels of the object. The objects are then equally
weighted, even when they have very different magnitudes, sizes and profiles.

The intrinsic ellipticity of the galaxies induces a statistical
dispersion on the shear estimate of $\sqrt{\bar \epsilon /N_g}$,
 where $\bar \epsilon$ is the mean ellipticity of $N_g$ galaxies.
Instrumental errors, tracking errors or anisotropic PSF may be removed
provided  they are measurable on the stellar profiles (Bonnet \& Mellier
1995, Kaiser et al. 1995).

The photon noise is a source of error of this method. Indeed,
the distortion is computed from one object, the ACF itself. Since the
noise polarizes randomly an object, a high noise level makes
 the measurement of the weak distortion impossible. This lead to the conclusion
that a given level of noise corresponds to a distortion threshold $\gamma_0$
below which the measured distortion is not reliable. We quantified this
threshold from simulations.

\section{Conclusion}

An optimum analysis of the lensing effects requires the measurement of both
the distortion and the magnification to comfirm and improve the results
quoted in Table (1), and to measure the very weak shear caused by large scale
structures.
Van Waerkebe et al. (1996) have proposed a new and independent method to 
measure the
gravitational distortion of the background galaxies
from the auto-correlation function of the brightness distribution.
It does not require any shape, size and centroid determination of individual
galaxies, and avoids possible systematics.
Moreover the resulting shear is unique and does not depends on the choice
of the detection criteria.

The method has been checked on simulated and true data (Q2345 and CL0024).
An example of the shear analysis using the ACF on simulated data
is shown on Figures 1,2.
The shear maps of the real images Q2345 and CL0024
were previously obtained by Bonnet 
et al. (1993,1994) with the standard method of individual galaxy analysis. Our
results are in very good agreement. Moreover, because of
the increase of the sensibility with our method, we predict the
existence of a new gravitational deflector in the field of Q2345.
Further observations will check this point. Because of its
simplicity and robustness, this method is
well adapted to measure weak shear caused by large scale structure for
which a large number of galaxies ($\sim 100000$) is required.

\acknowledgements
We thanks P. Schneider, F. Bernardeau and B. Fort for discussions and
enthusiastic support. LVW thanks B. Guiderdoni for his invitation
and the Moriond's staff for hospitality and financial support.

\section{References}

Bartelmann, M., Narayan, R. (1995) ApJ 451, 60.

{\parindent=0pt
Bonnet, H., Fort, B., Kneib, J-P., Mellier, Y., Soucail, G. (1993) A\&A
280, L7

Bonnet, H., Mellier, Y., Fort, B. (1994) ApJ 427, L83.

Bonnet, H., Mellier, Y. (1995) A\&A 303, 331.

Broadhurst, T. (1996) SISSA preprint astro-ph/9511150.

Broadhurst, T., Taylor, A.N., Peacock (1995) ApJ 438, 49.

Falhman, G., Kaiser, N., Squires, G., Woods, D. (1994) ApJ 437. 56.

Kaiser, N. (1996) SISSA preprint astro-ph/9509019.

Kaiser, N., Squires, G. (1993) ApJ 404, 441

Kaiser, N., Squires, G., Broadhurst, T. (1995) ApJ 449, 460.

Luppino, G., Kaiser, N. (1996) SISSA preprint astro-ph/9601194.

Miralda-Escud\'e, J., Babul, A. (1995) ApJ, 449, 18.

Navarro, J., Frenk, C., White, S. (1995) MNRAS, 275, 720.

Schneider, P., Ehlers, J., Falco, E. E., (1992), {\it Gravitational
Lenses}, Springer.

Schneider, P., Seitz, C. (1995) A\&A 294, 411.

Seitz, C., Kneib, J.P., Schneider, P., Seitz, S., (1996) in press. 

Smail, I., Ellis, R.S., Fitchett, M. (1994) MNRAS 270, 245.

Squires, G., Kaiser, N., Babul, A., Fahlmann, G., Woods, D., Neumann,
D.M., B\"ohringer, H. (1995) submitted.

Squires, G., Kaiser, N., Falhman, G., Babul, A., Woods, D. (1996) SISSA
preprint astro-ph/9602105.

Tyson, J.A., Fisher, P. (1995) ApJL, 349, L1.

Van Waerbeke, L., Mellier, Y., Schneider, P., Fort, B., Mathez, G.  (1996)
A\&A in press.
}

\begin{figure}
\caption{Simulation of a 4 hours exposure at CFHT in the B band on a
$3.5"\times 3.5"$ field. The seeing is $0.7"$ with
no tracking errors. Galaxies are lensed by an isothermal sphere
($\sigma=1000$km/s), with a core radius of $4''$ located at $200''$
bottom from the field center. The lens redshift is 0.17 and the mean
redshift of the sources is 1. The segments show the local orientation
of the shear. Their length is proportional to the shear intensity.}
\end{figure}

\begin{figure}
\psfig{figure=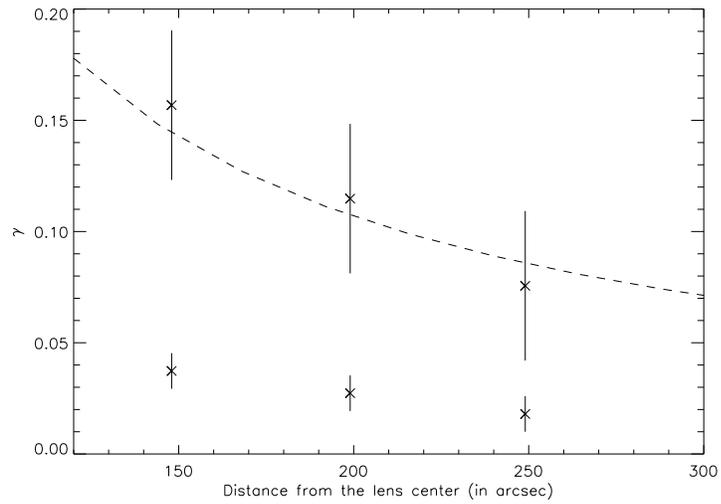,width=10. cm}
\caption{1-dimensional shear profile from the simulation of Fig.1.
At the bottom the uncalibrated measure points are drawn. The
theoretically expected shear profile is plotted as the dashed line.}
\end{figure}

\end{document}